\documentstyle[fleqn]{tp}

\input epsf

\def\etal{{\it et\thinspace al\/}}

\def\eg{{e.g.\ }}

\def\etal{{et al.\ }}

\def\mpc{\mbox{Mpc}}
\def\msun{{M_{\odot}}}

\begin{document}

\twocolumn[
\title{Semi-analytical galaxy formation models and the high redshift universe}
\author{Cedric Lacey$^1$, Carlton Baugh$^2$, Shaun Cole$^2$, Carlos
Frenk$^2$\\
  and Fabio Governato$^2$
\\
{\it $^1$Theoretical Astrophysics Center, Copenhagen, Denmark}\\
{\it $^2$Physics Department, Durham University, Durham, UK}}
\vspace*{16pt}   

ABSTRACT.\ 
Semi-analytical models of galaxy formation based on
hierarchical clustering now make a wide range of predictions for
observable properties of galaxies at low and high redshift. This
article concentrates on 2 aspects: (1)~Self-consistent modelling of
dust absorption predicts a mean UV extinction $A_{UV} \sim 1\ mag$,
depending only weakly on redshift, and similar to observational
estimates. (2)~The models predict that the Lyman-break galaxies found
at $z\sim 3$ should be strongly clustered with a comoving correlation
length $r_0 = 4-7\ \mpc/h$ (depending on the cosmology), in good
agreement with subsequent observational determinations.
\endabstract]

\markboth{C. Lacey et al.}{Semi-analytical galaxy formation models}

\small

\section{Introduction}
The technique of semi-analytical modelling is a powerful tool for
making predictions about the observable properties of galaxies at low
and high redshift, based on hierarchical clustering models of
structure formation, and thus for testing ideas about galaxy
formation. The models we describe here are a greatly improved version
of those of Cole \etal (1994), and are fully described in Cole \etal
(1998). The models include the following processes: (a) formation of
dark halos through merging; (b) shock-heating and radiative cooling of
gas within halos; (c) collapse of cool gas to rotationally-supported
disks; (d) star formation from cold gas; (e) feedback from supernovae;
(f) chemical evolution; (g) galaxy mergers producing elliptical
galaxies and bulges; (h) luminosity evolution of stellar populations;
and (i) absorption by dust. The treatment of dust absorption is new,
and is described in Section~\ref{dust}. Predictions for galaxy
clustering at high redshift are described in
Section~\ref{cluster}. 

The outputs from the models are masses, luminosities, colours, sizes,
morphologies, circular velocities and chemical compositions for the
whole galaxy population at any redshift. The models depend on several
parameters relating to star formation and feedback, which are chosen
to match observations of present-day galaxies, as described in Cole
\etal (1998). Once normalized in this way, the models match most
observations of present-day galaxies well, and can be used to make
predictions for the evolution of galaxy properties with redshift.

\begin{figure*}
  \centering\mbox{\epsfxsize=9.truecm \epsfysize=6.truecm 
\epsfbox[30 180 580 650]{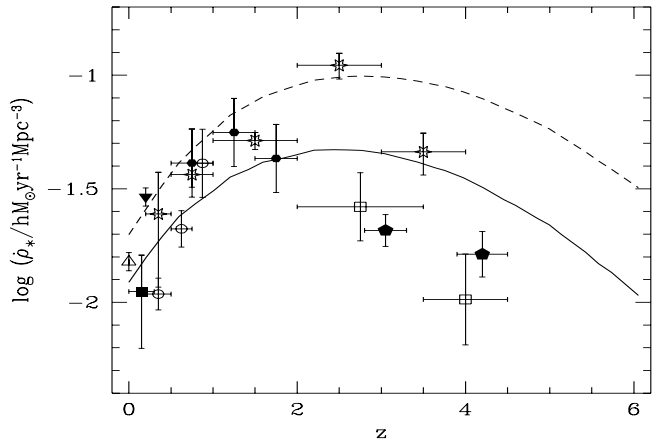}}
\caption[]{The cosmic star formation history. The curves show the
model predictions, for a CDM model with $\Omega_0=0.3$,
$\Lambda_0=0.7$ and a Kennicutt (1983) IMF. The $1500\AA$ luminosity
density from the model has been converted to a star formation density
using a constant conversion factor, as for the observations. The
dashed curve shows the model without dust, and the solid curve the
model prediction including dust. The symbols show observational
estimates of the star formation rate per comoving volume, derived from
$H\alpha$ or UV luminosity densities assuming a Kennicutt IMF and
solar metallicity. The triangles are from $H\alpha$ luminosities, and
have been corrected observationally for dust (and so should be
compared to the dashed curve), while the other points are from
rest-frame UV ($1500 - 3000 \AA$) luminosities, and do not include any
dust correction (and should be compared to the solid curve). The
references for the observational data are: open triangle: Gallego
\etal (1995); filled triangle: Tresse \& Maddox (1998); filled square:
Treyer \etal (1998); open circles: Lilly \etal (1996); stars: Sawicki
\etal (1997); filled hexagons: Conolly \etal (1997); open squares:
Madau \etal (1998). Filled pentagons show preliminary results
from a new survey by Steidel \etal (1998), and may be revised when the
survey is completed.}
\label{fig:1}
\end{figure*}

\section{Dust absorption}\label{dust}
Absorption of starlight by dust has a significant effect on optical
luminosities and colours of galaxies, and a large effect on the far-UV
luminosities which are used as the main tracer of star formation rates
at high redshift. We calculate dust absorption in a self-consistent
way for each galaxy, using the $3D$ radiative transfer models of
Ferrara \etal (1998). The latter models include stars in a bulge and
exponential disk, and dust in an exponential disk, and give the net
attenuation of galaxy light as a function of wavelength, inclination,
and the galaxy parameters. These models are a major improvement over
previous treatments, which assume a $1D$ screen or slab geometry for
the stars and dust. The most important parameter in the dust models is
$\tau_0$, the central face-on optical depth in dust (\eg in the
$V$-band). This is calculated directly from the semi-analytical model,
as:
\begin{equation}
\tau_0 \propto {M_{dust} \over \pi r_{disk}^2} \propto {M_{gas}Z_{gas}
\over r_{disk}^2}
\end{equation}
The dust is assumed to have a solar neighbourhood extinction law, with
the dust/gas ratio scaling as the gas metallicity $Z_{gas}$. $Z_{gas}$
is obtained from our chemical evolution calculation, and the disk
radius $r_{disk}$ is calculated based on angular momentum
conservation. We assume a ratio of vertical scaleheights
$h_z(dust)/h_z(stars)=1$, but the results are only weakly dependent on
this value. Thus, our calculation of dust absorption has essentially
no free parameters.

We find in our models that the mean dust extinction increases strongly
with galaxy luminosity, this being an effect mainly of the increase of
mean surface density with galaxy mass. On the other hand, the
dependence of net extinction on wavelength is much weaker than
predicted by a simple foreground screen model for the
dust. Figure~\ref{fig:1} shows the effects of dust on the inferred
star formation history of the universe, for a CDM model with
$\Omega_0=0.3$, $\Lambda_0=0.7$ and a Kennicutt (1983) IMF. Since most
observational estimates of the star formation density are derived from
far-UV luminosities, we have followed the same approach for the
models, taking the predicted $1500$\AA\ luminosity density (with or
without dust) and converting it to an SFR density assuming the same
IMF and a constant (solar) metallicity. The mean $1500$\AA\ extinction
is predicted to vary only slowly with redshift, increasing from $0.5\
mag$ at $z=0$ to $1.2\ mag$ at $z=6$, the smaller dust content of
higher redshift galaxies being more than compensated by their smaller
sizes, resulting in higher optical depths. The mean extinction at
$3000$\AA\ is similar to that at $1500$\AA. These values for the UV
extinction are similar to those estimated observationally for high
redshift galaxies \eg Pettini \etal (1998).

The model shown predicts that the star formation rate per comoving
volume peaks at $z\sim 3$. The model prediction appears to be broadly
consistent with the current observational data, but, given the large
error bars and significant scatter between different observations, a
definitive comparison is not yet possible.

\begin{figure*}
  \centering\mbox{\epsfxsize=9.truecm \epsfysize=6.truecm 
\epsfbox[60 270 570 670]{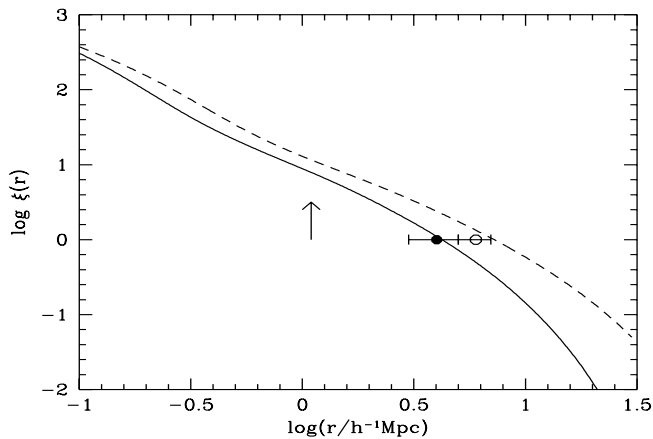}}
\caption[]{The clustering of Lyman-break galaxies brighter than
$R_{AB}=25$ at $z\approx 3$, based on an analytical halo clustering
model. The solid line shows the two-point correlation function in
comoving coordinates for a CDM model with $\Omega=1$, while the dashed
line shows a CDM model with $\Omega_0=0.3$, $\Lambda_0=0.7$. The
symbols with error bars show the values of $r_0$ (defined by
$\xi(r_0)=1$) found observationally by Adelberger \etal (1998) for
different assumed cosmologies: filled symbol: $\Omega=1$; open symbol:
$\Omega_0=0.3$, $\Lambda_0=0.7$.}
\label{fig:2}
\end{figure*}

\begin{figure*}
  \centering\mbox{\epsfxsize=9.truecm  
\epsfbox[40 170 570 680]{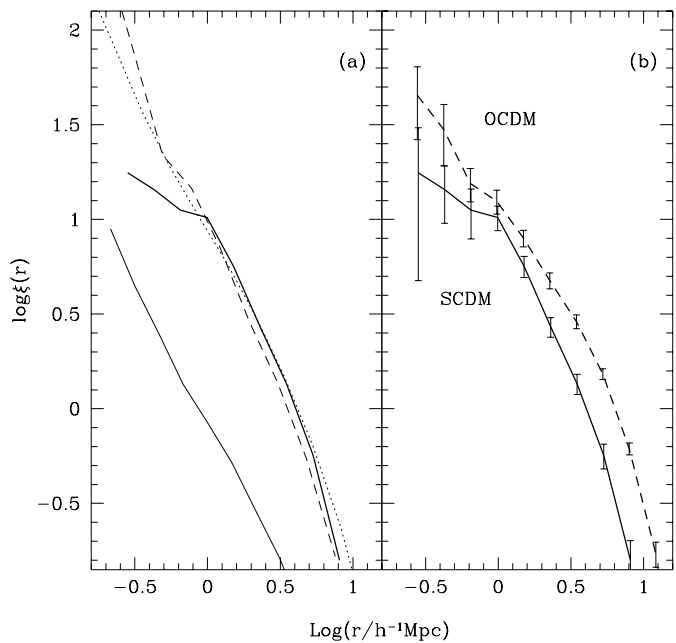}}
\caption[]{The clustering of Lyman-break galaxies at $z= 3$,
calculated using a combined N-body/semi-analytic method. (a)
Correlation functions in comoving coordinates for the $\Omega=1$ CDM
model. Lower solid line: dark matter correlation function; dashed
line: real space correlation function of Lyman-break galaxies; upper
solid line: redshift space correlation function of Lyman-break
galaxies; dotted line: real space correlation function of Lyman-break
galaxies calculated from the analytical model, as in
Figure~\ref{fig:2}. (b) Redshift space correlation functions for
$\Omega=1$ (SCDM) and $\Omega_0=0.3$, $\Lambda_0=0$ (OCDM) models.}
\label{fig:3}
\end{figure*}

\section{Clustering of high redshift galaxies}\label{cluster}
Semi-analytical galaxy formation models predict what should be the
galaxy populations in dark halos of different masses, and thus predict
what should be the {\em clustering bias} of galaxies relative to dark
matter. Measurements of the clustering of galaxies at different
redshifts and different luminosities thus provide an important test of
galaxy formation models. The clustering amplitude has now been
measured for a large sample of normal galaxies at $z\sim 3$, detected
through their Lyman break features (\eg Steidel, these
proceedings). The predictions of our semi-analytical model for the
properties of these Lyman-break galaxies originally found by Steidel
\etal (1996) have been presented in Baugh \etal (1998). They show that
CDM-based models with reasonable parameters are able to explain the
observed number density and other properties of these galaxies. An
important prediction of the models, made in advance of the
observations, was that the Lyman-break galaxies at $z\sim 3$ should be
strongly clustered. Figure~\ref{fig:2} shows the prediction of Baugh
\etal (1998) for the two-point correlation function of Lyman-break
galaxies brighter than apparent magnitude $R_{AB}=25$ at $z\approx 3$,
for two different cosmological models.

The Baugh \etal prediction was based on an analytical estimate
of the non-linear evolution of the power spectrum of density
fluctuations, and on the Mo \& White (1996) analytical formula for the
clustering bias $b$ of dark matter halos of mass $M$ at redshift $z$:
\begin{equation}
b(M,z) = 1 + {1\over \delta_c(z)}\left[ {\delta_c^2(z)\over \sigma^2(M,z)}
- 1 \right]
\label{eq:2}
\end{equation}
where $\delta_c(z)$ is the critical linear overdensity for an object
to collapse at redshift $z$, and $\sigma(M,z)$ is the variance of
linear density fluctuations of mass $M$. This formula for the halo
bias is expected to be valid down to scales somewhat smaller than the
comoving radii of the halos concerned (shown by an arrow in
Figure~\ref{fig:2}), at which point halo-halo exclusion effects become
significant. The semi-analytical models predict that the Lyman-break
galaxies found by Steidel \etal should lie in halos of masses $M\sim
10^{12} \msun$. This is much larger than the typical halo mass $M_*$
(defined by $\sigma(M_*,z)=\delta_c(z)$) at $z=3$, and so according to
equation~\ref{eq:2}, these halos and the galaxies they contain should
be highly biased. Baugh \etal calculated the mean clustering bias for
the $R_{AB}<25$ Lyman-break galaxies, and found $b\approx 4$ for both
$\Omega=1$ and $\Omega_0=0.3$, $\Lambda_0=0.7$ CDM models. The
accuracy of the analytical approximations used for the halo clustering
was subsequently confirmed by Governato \etal (1998), who combined the
semi-analytical models with N-body simulations, and found essentially
identical results for the clustering of Lyman break galaxies. This is
shown in Figure~\ref{fig:3}.

A useful measure of the clustering amplitude is the correlation length
$r_0$, defined to be the separation (in comoving coordinates) where
$\xi(r_0)=1$. The Baugh \etal prediction for the Lyman-break galaxies
was $r_0=4\ \mpc/h$ for $\Omega=1$ and $r_0=7\ \mpc/h$ for
$\Omega_0=0.3$, $\Lambda_0=0.7$. These predictions agree remarkably
well with the values subsequently measured by Adelberger \etal (1998)
from a counts-in-cells analysis of redshift surveys of these galaxies,
which were $r_0= (4\pm1)\ \mpc/h$ for $\Omega=1$ and $r_0=(6\pm1)\
\mpc/h$ for $\Omega_0=0.3$, $\Lambda_0=0.7$. These values are also
plotted in Figure~\ref{fig:2} for comparison with the models. This
successful prediction of the clustering of high-redshift galaxies is
strong evidence for these galaxies being in the most massive halos at
this redshift, and is an important confirmation of the validity of the
semi-analytical models.




\begin{thebibliography}{99}

\bibitem{ref:1} Adelberger, K., Steidel, C., Giavalisco, M.,
Dickinson, M., Pettini, M., Kellog, M., 1998, ApJ, 505, 18

\bibitem{ref:1} Baugh, C., Cole, S., Frenk, C., Lacey, C., 1998, ApJ,
498, 504

\bibitem{ref:1} Cole, S., Aragon-Salamanca, A., Frenk, C., Navarro,
J., Zepf, S., 1994, MNRAS, 271, 281

\bibitem{ref:1} Cole, S., Lacey, C., Baugh, C., Frenk, C., 1998, in
preparation 

\bibitem{ref:1} Conolly, A., Szalay, A., Dickinson, M., SubbaRao, M.,
Brunner, R., 1997, ApJ, 486, L11

\bibitem{ref:1} Ferrara, A., Bianchi, S., Cimatti, A., Giovanardi,
C. 1998, in preparation.

\bibitem{ref:1} Gallego, J., Zamorano, J., Aragon-Salamanca, A., Rego,
M., 1995, ApJ, 455, L1

\bibitem{ref:1} Governato, F., Baugh, C., Frenk, C., Cole, S., Lacey,
C., Quinn, T., Stadel, J., 1998, Nature, 392, 359

\bibitem{ref:1} Kennicutt, R., 1983, ApJ, 272, 54

\bibitem{ref:1} Lilly, S., LeFevre, O., Hammer, F., Crampton, D.,
1996, ApJ, 460, L1

\bibitem{ref:1} Madau, P., Pozzetti, L., Dickinson, M., 1998, ApJ,
498, 106

\bibitem{ref:1} Mo, H., White, S., 1996, MNRAS, 282, 347

\bibitem{ref:1} Pettini, M., Kellog, M., Steidel, C., Dickinson, M.,
Adelberger, K., Giavalisco, M., 1998, ApJ, in press; astro-ph/9806219

\bibitem{ref:1} Sawicki, M., Lin, H., Yee, H., 1997, AJ, 113, 1

\bibitem{ref:1} Steidel, C., Giavalisco, M., Pettini, M., Dickinson, M.,
Adelberger, K., 1998, ApJ, 462, L17

\bibitem{ref:1} Tresse, L., Maddox, S., 1998, ApJ, 495, 691

\bibitem{ref:1} Treyer, M., Ellis, R., Milliard, B., Donas, J.,
Bridges, T., 1998, MNRAS, 300, 303

\end{thebibliography}
\end{document}